\documentclass[aps,prl,showpacs,twocolumn,groupedaddress,floatfix,amsfonts]{revtex4}

\usepackage{graphicx}  
\usepackage{dcolumn}   
\usepackage{bm}        
\usepackage{amssymb}   
\usepackage{setspace}  

\newcommand{\simleq}{\; \raisebox{-0.4ex}{\tiny$\stackrel
{{\textstyle<}}{\sim}$}\;}
\def\lsim{\mathrel{\rlap{\lower4pt\hbox{\hskip1pt$\sim$}}\raise1pt\hbox{$<$}}}
\def\gsim{\mathrel{\rlap{\lower4pt\hbox{\hskip1pt$\sim$}}\raise1pt\hbox{$>$}}}
\def\MET{{\mbox{$E\kern-0.57em\raise0.19ex\hbox{/}_{T}~$}}}
\def\METnoSpace{{\mbox{$E\kern-0.57em\raise0.19ex\hbox{/}_{T}$}}}

\begin{document}








\title{Search for a resonance decaying into $WZ$ boson pairs in $p\bar{p}$ collisions} 

%
\author{V.M.~Abazov$^{37}$}
\author{B.~Abbott$^{75}$}
\author{M.~Abolins$^{65}$}
\author{B.S.~Acharya$^{30}$}
\author{M.~Adams$^{51}$}
\author{T.~Adams$^{49}$}
\author{E.~Aguilo$^{6}$}
\author{M.~Ahsan$^{59}$}
\author{G.D.~Alexeev$^{37}$}
\author{G.~Alkhazov$^{41}$}
% ...

\affiliation{\vspace{0.1 in}(The D\O\ Collaboration)\vspace{0.1 in}}
% ...
\affiliation{$^{6}$University of Alberta, Edmonton, Alberta, Canada;
                Simon Fraser University, Burnaby, British Columbia, Canada;
                York University, Toronto, Ontario, Canada and
                McGill University, Montreal, Quebec, Canada}
% ...
\affiliation{$^{30}$Tata Institute of Fundamental Research, Mumbai, India}
% ...
\affiliation{$^{37}$Joint Institute for Nuclear Research, Dubna, Russia}
% ...
\affiliation{$^{41}$Petersburg Nuclear Physics Institute,
                St. Petersburg, Russia}
% ...
\affiliation{$^{49}$Florida State University, Tallahassee, Florida 32306, USA}
% ...
\affiliation{$^{51}$University of Illinois at Chicago,
                Chicago, Illinois 60607, USA}
% ...
\affiliation{$^{59}$Kansas State University, Manhattan, Kansas 66506, USA}
% ...
\affiliation{$^{65}$Michigan State University,
                East Lansing, Michigan 48824, USA}
% ...
\affiliation{$^{75}$University of Oklahoma, Norman, Oklahoma 73019, USA}
% ...


\begin{abstract}
We present the first search for an electrically charged resonance $W^\prime$
decaying to a $WZ$ boson pair using 4.1~fb$^{-1}$ of integrated luminosity
collected with the D0 detector at the Fermilab Tevatron $p\bar{p}$ collider.
The $WZ$ pairs are reconstructed through their decays into three 
charged leptons ($\ell=e,\mu$). A total of 9 data events is observed 
in good agreement with the background prediction.
We set 95\% C.L. limits on the $W^\prime WZ$ coupling and on the 
$W^\prime$ production cross section multiplied by the branching fractions.
We also exclude $W^\prime$ masses between 188 and 520 GeV within a
simple extension of the standard model and set the most restrictive limits
to date on low-scale technicolor models.

\end{abstract}

\maketitle


The standard model (SM) of particle physics is widely believed to be a low energy 
approximation of a more fundamental theory of elementary particles and their interactions.
Many extensions of the SM, such as the sequential standard model (SSM)~\cite{ssm}, 
extra dimensions~\cite{extradim}, little Higgs~\cite{littleHiggs}, and 
technicolor~\cite{LSTC} models, predict new heavy $W^\prime$ resonances decaying to
a pair of electroweak $W$ and $Z$ bosons. Some models~\cite{extradim, littleHiggs, LSTC} also offer
an alternative to the SM mechanism of electroweak symmetry breaking.
Thus, the observation of resonant $WZ$ boson production would not only manifest 
new physics beyond the SM, but also could yield an insight into the origin of mass.

This Letter describes the first search for a heavy charged boson, 
referred to as the $W^\prime$, decaying to $W$ and $Z$ bosons.
The CDF and D0 collaborations have searched for a $W^\prime$ decaying to 
fermions~\cite{CDFrun2wp1,CDFrun2wp2,run2wp3}. 
Current limits exclude $W^\prime$ with masses $\simleq$~1~TeV at 95\%~C.L.,
assuming the SSM as benchmark and that the $W^\prime \to WZ$ decay is fully suppressed.
Thus, our search is complementary to the previous studies. 

In technicolor, particles such as $\rho_T$ and $a_T$ have narrow widths and can decay to
$WZ$ bosons. The experimental signature of these particles is therefore similar to that of a $W^\prime$.
We will interpret the results of our search within the low-scale technicolor model (LSTC),
where the masses of $\rho_T$ and $a_T$ are predicted to be below 500~GeV,
well within the energy reach of the Tevatron. Since $\rho_T$ and $a_T$ have almost the same
mass we refer to them collectively as $\rho_T$.
The branching fraction for $\rho_T\to WZ$ depends 
strongly on the relative masses of the technipion, $M(\pi_T)$, and technirho,
$M(\rho_T)$.
The D0 collaboration searched previously for technicolor in the $W\pi_T\to W+{\rm jets}$ final 
state~\cite{run2wp4}, which is one of the major decay channels for light technipions. 
In this Letter we present a search in a previously unexplored region of LSTC phase space
with $M(\pi_T) \simleq M(\rho_T)$ where $\rho_T$ decays predominantly to a $WZ$ boson pair.

We perform the search using data collected with the D0 detector~\cite{run2det}
at the Fermilab Tevatron $p\bar{p}$ collider at a center of mass energy 
of $\sqrt{s}=1.96$~TeV. After applying data quality and trigger requirements, the 
integrated luminosity corresponds to 4.1~fb$^{-1}$.

The Monte Carlo (MC) samples for resonant $WZ$ signal and SM backgrounds
are generated using {\sc pythia}~\cite{pythia}, with the exception of $Z+{\rm jets}$
and $t\bar{t}$ processes that are generated using {\sc alpgen}~\cite{alpgen} 
interfaced with {\sc pythia} for showering and hadronization. 
All MC samples are passed through a full {\sc geant}~\cite{geant} simulation of the D0 detector. 
The MC is corrected to describe the luminosity dependence of the trigger 
and reconstruction efficiencies in data and the contribution from multiple $p\bar p$ interactions. 
The MC sample for signal is produced assuming SSM $W^\prime$ production for masses 
starting at 180, 190, 200~GeV and then up to 1~TeV in steps of 50~GeV, 
using CTEQ6L1~\cite{cteq} parton distribution functions (PDF). 
The interference between signal and the SM $s$-channel $WZ$ 
production~\cite{rizzo} is negligible and is not taken into account. 
We generate technicolor $WZ$ samples using typical parametrization of the 
LSTC phenomenology implemented in {\sc pythia}~\cite{StrawManTC}
to estimate the leading order cross section, efficiency, and acceptance of
the selection criteria of the $\rho_T\to WZ$ production.
All MC samples are normalized to the integrated luminosity using next-to-leading order 
cross section calculations, with the exception of the $W^\prime$ signal cross section, which is known 
to next-to-next-to-leading order (NNLO)~\cite{wprime_nnlo}. 
All MC samples are subject to the same event selection as applied to data.

In this search we select events where both the $W$ and the $Z$ bosons decay
leptonically and consider only final states with electrons and muons. 
Candidate events with at least two final state electrons are selected 
using single-electron triggers, while those with at least two muons 
are selected using single-muon triggers resulting in efficiencies of 100\%
and 92\% respectively for signal events.
The events are required to have missing transverse energy $\METnoSpace>30$~GeV~\cite{delta_r} 
(from the undetected neutrino) and at least three charged leptons with
transverse momenta $p_T > 20$~GeV satisfying the electron or muon
identification criteria described below.
An electron candidate is identified as a central track matched to an isolated cluster of
energy in the calorimeter, with a shower shape consistent with that of an electron, in 
the pseudorapidity range $|\eta|<1.1$ or  $1.5<|\eta|<2.5$.  
A muon candidate is reconstructed from segments in the muon spectrometer 
matched to a central track, and is required to be within  $|\eta|<2$. 
The muon candidate must be isolated from other activity in the tracker and calorimeter.

The selection of $WZ$ candidate events is done in two steps. We first require the presence of
a candidate $Z$ boson by selecting the electron pairs and muon pairs with opposite
electric charges that have invariant mass nearest to the mass of the $Z$ boson. 
The reconstructed mass of the $Z$ boson candidate must be between 80 and 102 GeV for an 
electron pair and between 70 and 110 GeV for a muon pair. 
Then, we select the highest transverse momentum lepton among the remaining lepton
candidates in the event as the lepton from the $W$ boson decay.
The $W$ and $Z$ bosons produced from heavy resonances can be 
highly boosted, resulting in a large spatial separation between leptons 
from the $W$ and $Z$ decays. To reduce background, we require the lepton from 
the $W$ boson decay to be separated by 
$\Delta{\cal R} = \sqrt{(\Delta{\cal \eta})^2+(\Delta{\cal \phi})^2} > 1.2$
from $Z$ decay leptons.
	
Several background processes contribute to the trilepton + $\METnoSpace$ final state.
The largest background having at least three genuine leptons in the final state is
from SM $WZ$ production, followed by the $ZZ$ process, where one of the leptons from 
the $Z$ boson is not reconstructed and gives rise to $\METnoSpace$. These are estimated
from MC simulation. The instrumental background is due to misidentification
of a lepton in processes such as $Z+{\rm jets}$, $Z\gamma$, and $t\bar{t}$.  
Contribution from $t\bar{t}$ is estimated from MC simulation and found to be 
negligible. 
$Z+{\rm jets}$ and $Z\gamma$ productions are the major instrumental backgrounds
and they are estimated using data driven techniques described below.

Jets from $Z+{\rm jets}$ production can be misidentified as either
an electron or a muon from $W$ boson decay. 
To estimate this contribution, we select a sample of $Z$ boson decays with an
additional {\it "false"} lepton candidate for each final state. For the $Z$+electron
final state the lepton candidate is required to have most of its energy
deposited in the electromagnetic calorimeter and satisfy the electron isolation criteria, but
at the same time a shower shape inconsistent with that of an electron. For the $Z$+muon
final state, the lepton candidate is required to fail the isolation criteria used to select
muons. These requirements ensure that the lepton is either a misidentified jet or a 
lepton from a semileptonic decay of a heavy-flavor quark.
The contribution from the $Z+{\rm jets}$ background with misidentified leptons is
estimated by scaling the number of events in this sample with a $p_T$-dependent
ratio of misidentified leptons passing the two different sets of criteria measured
in a multijet data sample depleted of true isolated leptons.

The channels with $W\rightarrow e\nu$ decays can be mimicked by the initial or 
final state radiation $Z\gamma$ processes where a photon is either
incorrectly matched to a track, or converts, and one of the conversion electrons 
is selected as the candidate for $W$ boson decay. To estimate the contribution 
from this background, we measure the rate at which a photon can be misidentified as 
an electron in $Z\to\mu\mu\gamma$ final states in data,
as it offers a virtually background-free source of photons 
because of the $\mu\mu\gamma$ invariant mass constraint to the $M(Z)$. 
We choose the muon decay of the $Z$ boson to avoid ambiguity in assigning
the electromagnetic showers in the $ee\gamma$ final states.
The $Z\gamma$ contribution is estimated by folding in the $p_{T}$-dependent 
photon to electron misidentification rate with the $p_T$ 
distribution of $\gamma$ in the $Z\gamma$ Monte Carlo simulation~\cite{Baur}.

The selection criteria yield 9 events in data with an estimate of 10.2~$\pm$~1.6 background events.
The final numbers of observed candidates and expected signal and background events
with total uncertainties are summarized in 
Table~\ref{tab:results}. The expected and observed yields for the four 
independent samples used in this search are given in Table~\ref{tab:results_fourCh}.
Several sources of systematic uncertainties are considered here. 
The major systematic uncertainty is associated with the modeling of the trigger,
the lepton identification efficiency and the detector acceptance.
It is estimated to be 15\%.
This uncertainty is taken as fully correlated between signal and background.
We assign to the $Z\gamma$ background estimation a systematic uncertainty 
of 100\% for any potential mis-modeling of $\METnoSpace$. 
The dominant systematic uncertainty on $Z+{\rm jets}$ background is from the limited statistics
of the $Z+{\it "false"}$ lepton sample. We estimate this uncertainty to be 40\%.
Finally, the uncertainty on integrated luminosity is 6.1\%~\cite{lumi}, and the uncertainty on the 
theoretical NNLO production cross section of signal is 5\%. 

\begin{table}[ht]
\centering
\small
\begin{tabular}{lc} \hline\hline
Source               & Total \\ \hline
$W^\prime$(500 GeV)  & $4.4 \pm 1.1$   \\ \hline
$WZ$                 & $9.0 \pm 1.5$   \\
$ZZ$                 & $1.0 \pm 0.2$  \\
$Z+{\rm jets}$ 	     & $0.2 \pm 0.1$   \\ 
$Z\gamma$            & $0.1 \pm 0.1$ \\ 
Total                & $10.2 \pm 1.6$  \\ \hline
Observed             & 9                 \\ \hline\hline
\end{tabular}
\caption{Number of data events, expected number of signal events for a SSM $W^\prime$
	mass of 500~GeV and expected number of background events with statistical
	and systematic uncertainties.}
\label{tab:results}
\end{table}

\begin{table*} [ht]
\centering
\small
\begin{tabular}{lccccccc} 
\hline \hline
Mode & $WZ$ & $ZZ$ & $Z+{\rm jets}$ & $Z\gamma$ & Total & $W^\prime$ & Data \\ \hline
eee             & $1.4 \pm 0.3$ & $0.07 \pm 0.02$ & $0.02 \pm 0.01$ & $0.03 \pm 0.03$ & $1.52 \pm 0.33$ & $1.07 \pm 0.28$ & 3 \\
ee$\mu$         & $2.0 \pm 0.4$ & $0.24 \pm 0.06$ & $0.07 \pm 0.04$ &         $< 0.01$     & $2.31 \pm 0.49$ & $1.17 \pm 0.31$& 2\\
e$\mu\mu$     & $2.0 \pm 0.4$ & $0.10 \pm 0.03$ & $0.04 \pm 0.02$ & $0.07 \pm 0.07$ & $2.21 \pm 0.46$ & $0.83 \pm 0.22$ & 2\\
$\mu\mu\mu$ & $3.6 \pm 0.8$ & $0.54 \pm 0.12$ & $0.05 \pm 0.03$ &        $< 0.01$      & $4.19 \pm 0.89$ & $1.28 \pm 0.34$ & 2\\ \hline\hline
\end{tabular}
\caption{Background estimation from the leading sources, the total background, expected signal, and observed events for each signature. 
         The signal corresponds to a SSM $W^\prime$ with a mass of 500~GeV. The uncertainties reflect both the statistics
	 of the MC and data samples and systematics.}	
\label{tab:results_fourCh}
\end{table*}

As the number of observed candidates is consistent with the background-only hypothesis,
we set limits on $W^\prime$ production in a modified frequentist approach~\cite{collie}
that uses a log-likelihood ratio ($LLR$) test statistic~\cite{llr}. 
It calculates the confidence levels for the signal + background, $CL_{s+b}$,
and background-only hypothesis, $CL_b$, by integrating the $LLR$ distributions obtained
from simulated pseudo-experiments using Poisson statistics. Systematic uncertainties are treated as 
uncertainties on the expected number of signal and background events. This ensures that the 
uncertainties and  their correlations are propagated to the outcome with proper weights. The 95\% confidence 
level (C.L.) limit on the cross section is then defined as a cross section for which the ratio 
$CL_s = CL_{s+b}/CL_{b}$ is 0.05. 

\begin{figure}
\begin{center}
\includegraphics[scale=0.43]{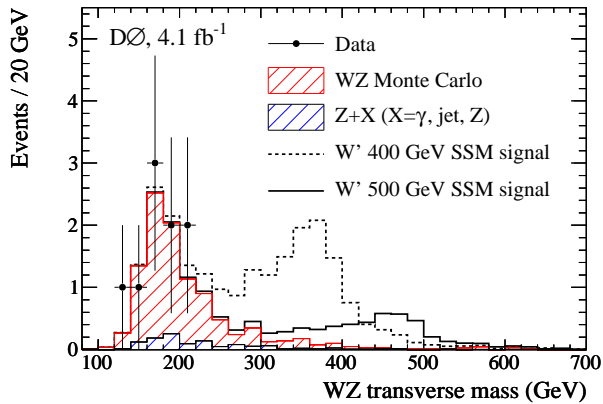}
\caption{Transverse mass distribution of the $WZ$ system in data with the major SM
	backgrounds and two SSM $W^\prime$ mass hypotheses overlaid (color online).
\label{fig:WZtrm}}
\end{center}
\end{figure}

\begin{figure}
\begin{center}
\includegraphics[scale=0.43]{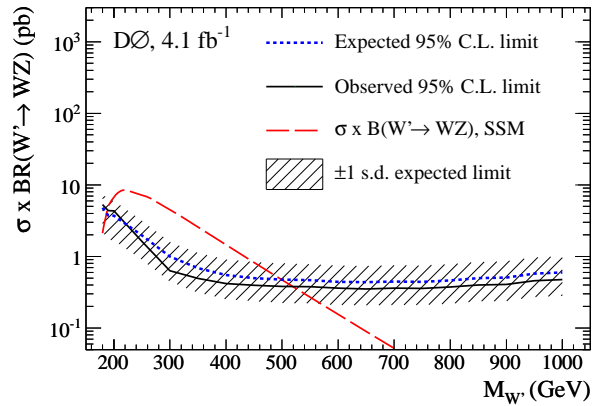}
\caption{ Observed and expected 95\% C.L. upper limits and $\pm 1$~s.d. band around
         the expected limits on the cross section multiplied by 
         $B(W^\prime\to WZ)$ with the SSM prediction overlaid (color online).
\label{fig:Limits}}
\end{center}
\end{figure}

We use the $WZ$ transverse mass to discriminate between the 
$W^\prime$ signal and the backgrounds in the limit setting procedure.
It is calculated as
\[ M_{T}=\sqrt{(E_T^Z+E_T^W)^2-(p_x^Z+p_x^W)^2-(p_y^Z+p_y^W)^2},\]
where $E_T^Z$ and $E_T^W$ are the scalar sums of the transverse momenta of the decay products of 
the $Z$ and $W$ candidates, respectively; while $p_x^Z$, $p_x^W$, $p_y^Z$, and $p_y^W$
are obtained by summing the $x$ and $y$ components of momenta of the respective decay particles.
In these sums, the transverse momentum of the neutrino arising from the $W$ boson decay is inferred from the 
direction and magnitude of $\METnoSpace$. 
The distribution of the $WZ$ transverse mass is given in Fig.~\ref{fig:WZtrm} for data, 
backgrounds, and two signal hypotheses. 
We obtain a limit on the production cross section of $W^\prime$ multiplied by the branching ratio
$B(W^\prime \to WZ)$ as a function of the $M(W^\prime)$ as shown in Fig.~\ref{fig:Limits}. 
This is the first limit to date on resonant $W^\prime\to WZ$ production.
Assuming SSM production, we exclude a $W^\prime$ with mass $188<M(W^\prime)<520$~GeV at 95\% C.L.. 
This result agrees with the expected sensitivity limit of $188 < M(W^\prime) < 497$~GeV. 

\begin{figure}
\begin{center}
\includegraphics[scale=0.43]{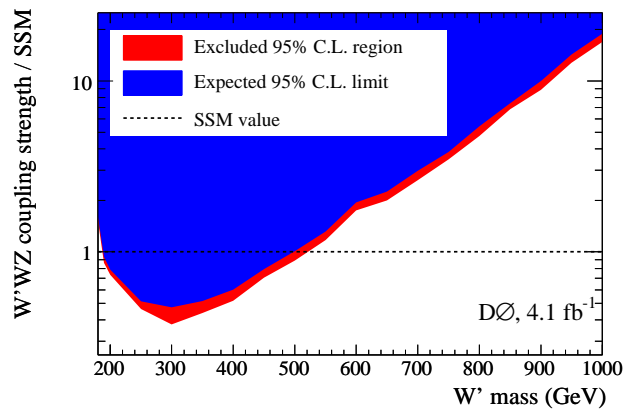}
\caption{Expected and excluded area of the $W^\prime WZ$ coupling strength normalized to the SSM value
as a function of the $W^\prime$ mass (color online).
\label{fig:LimitsWprWZ}}
\end{center}
\end{figure}

We also study the sensitivity to other models that predict a $W^\prime$-like resonance with width 
greater than in the SSM by varying the width of the $W^\prime$ resonance while keeping 
$\sigma(W^\prime)\times B(W^\prime\to WZ)$ fixed to the SSM value.
We find that the limits slightly degrade but stay within 1~standard deviation (s.d.) around the 
expected sensitivity limits for models with widths up to 25\% of the resonance mass. 
Since the limits have a limited sensitivity to the width of the $W^\prime$, we can exclude
more general models that predict $W^\prime$ bosons with arbitrary couplings to the $W$ and $Z$ bosons.
We interpret the results in terms of the $W^\prime WZ$ trilinear coupling normalized
to the SSM value as function of the $W^\prime$ mass (see Fig.~\ref{fig:LimitsWprWZ}).

\begin{figure}
\begin{center}
\includegraphics[scale=0.43]{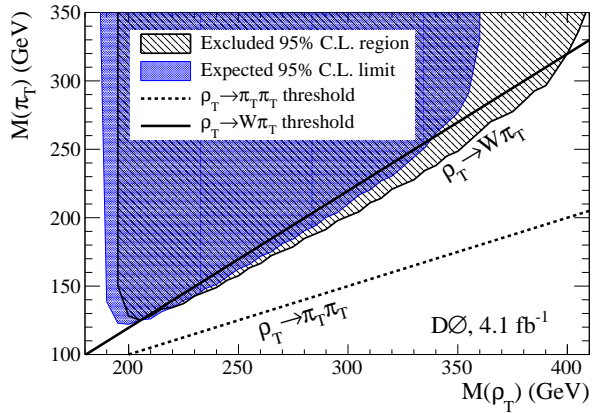}
\caption{Expected and excluded areas of the $\pi_T$ {\it vs.} $\rho_T$ masses
are given with the thresholds of the $\rho_T \to W \pi_T$
and $\rho_T \to \pi_T \pi_T$ overlaid (color online).
\label{fig:LimitsTC}}
\end{center}
\end{figure}

The limits on the resonant $WZ$ production cross section $\sigma\times B(W^\prime\to WZ)$ yield stringent 
constraints on the LSTC and exclude most of the allowed phase space where 
$\rho_T\to WZ$ decay is dominant.
The excluded and expected limits at 95\% C.L., as a function of the 
$\rho_T$ and $\pi_T$ masses, are shown in Fig.~\ref{fig:LimitsTC}. 

In summary, we have presented a search for hypothetical $W^\prime$ particles 
decaying to a pair of $WZ$ bosons using 
leptonic $W$ and $Z$ decay modes in 4.1 fb$^{-1}$ of Tevatron Run II data. 
We observe no evidence of resonant 
$WZ$ production, and set limits on the production cross section $\sigma \times B(W^\prime \to WZ)$. 
Within the SSM we exclude $W^\prime$ masses between 188 and 520~GeV at 95\% C.L. 
This is the best limit to date on $W^\prime\to WZ$ production and is complementary to previous 
searches~\cite{CDFrun2wp1,CDFrun2wp2,run2wp3} for $W^\prime$ decay to fermions. 
These limits are less stringent for the models that predict $W^\prime$ with width greater than that 
predicted by the SSM model, but stay within the 1 s.d. band around the expected SSM limits
for widths below 25\% of the $W^\prime$ mass.
The original limits are also interpreted within the technicolor model. We exclude $\rho_T$ with mass
between 208 and 408~GeV at 95\%~C.L. for $M(\rho_T) < M(\pi_T) + M(W)$. These are the most stringent constraints on a  typical LSTC phenomenology model~\cite{StrawManTC} when $\rho_T$ decays predominantly to $WZ$ boson pair.

%
We thank Kenneth Lane for useful discussions and help with interpretation of the results
within the TCSM parameter space and we thank the staffs at Fermilab and collaborating institutions, 
and acknowledge support from the 
DOE and NSF (USA);
CEA and CNRS/IN2P3 (France);
FASI, Rosatom and RFBR (Russia);
CNPq, FAPERJ, FAPESP and FUNDUNESP (Brazil);
DAE and DST (India);
Colciencias (Colombia);
CONACyT (Mexico);
KRF and KOSEF (Korea);
CONICET and UBACyT (Argentina);
FOM (The Netherlands);
STFC and the Royal Society (United Kingdom);
MSMT and GACR (Czech Republic);
CRC Program, CFI, NSERC and WestGrid Project (Canada);
BMBF and DFG (Germany);
SFI (Ireland);
The Swedish Research Council (Sweden);
CAS and CNSF (China);
and the
Alexander von Humboldt Foundation (Germany).


\begin{thebibliography}{99}
\bibitem{delta_r} The D0 detector utilizes a right-handed coordinate system with the $z$ axis
pointing in the direction of the proton beam and the $y$ axis pointing upwards. The azimuthal angle 
$\phi$ is defined in the $xy$ plane measured from the $x$ axis. The pseudorapidity is defined as
$\eta = -\rm{ln}[\rm{tan}(\theta/2)]$, where $\theta = \arctan(\sqrt{x^2+y^2}/z)$.
The transverse variables are defined as projections of the variables onto the $x-y$ plane. 
The missing transverse energy is the imbalance of the momentum estimated from the 
calorimeter and reconstructed muons in the $x-y$ plane.
\end{thebibliography}
\end{document}